\documentclass[journal]{IEEEtran}
\usepackage{amsmath,amsfonts}
\usepackage{lipsum,multicol}
\usepackage{algorithm}
\usepackage{bm}
\usepackage{array}
\usepackage{mathtools}
\usepackage{balance}
\usepackage{graphicx}

\def\BibTeX{{\rm B\kern-.05em{\sc i\kern-.025em b}\kern-.08em
T\kern-.1667em\lower.7ex\hbox{E}\kern-.125emX}}
\usepackage{balance}
\usepackage{multirow}
\usepackage{cite}

\usepackage[utf8]{inputenc}
 
\usepackage{pgfplots}
\usepackage{breqn}
\usepackage{stackengine}
\usepackage{mathtools,amssymb,lipsum}
\usepackage{cuted}

\pgfplotsset{compat=1.14}

\DeclareMathOperator{\var}{{var}}

\DeclareMathOperator{\E}{\mathbb{E}}

\DeclareMathOperator{\En}{\mathcal{E}}
\DeclareMathOperator{\Norm}{\mathcal{N}}
\DeclareMathOperator{\T}{\text{T}}
\DeclareMathOperator{\sign}{sign}

\DeclareMathOperator{\B}{\textbf{B}}
\DeclareMathOperator{\I}{\textbf{I}}
\DeclareMathOperator{\Q}{\textbf{Q}}
\DeclareMathOperator{\R}{\textbf{R}}
\DeclareMathOperator{\SM}{\textbf{S}}
\DeclareMathOperator{\U}{\textbf{U}}
\DeclareMathOperator{\V}{\textbf{V}}
\DeclareMathOperator{\W}{\textbf{W}}
\DeclareMathOperator{\Id}{\textbf{I}}

\DeclareMathOperator{\s}{\textbf{s}}
\DeclareMathOperator{\rv}{\textbf{r}}
\DeclareMathOperator{\x}{\textbf{x}}
\DeclareMathOperator{\y}{\textbf{y}}
\DeclareMathOperator{\dv}{\textbf{d}}
\DeclareMathOperator{\uv}{\textbf{u}}
\DeclareMathOperator{\Vb}{\textbf{V}}

\makeatletter
\newcommand{\@giventhatstar}[2]{\left[#1\;\middle|\;#2\right]}
\newcommand{\@giventhatnostar}[3][]{#1[#2\;#1|\;#3#1]}
\newcommand{\giventhat}{\@ifstar\@giventhatstar\@giventhatnostar}
\makeatother

\begin{document}
\title{Theoretical Analysis of Multi-Coding\\ with Arbitrary Correlations Among the Codes}
\author{Brian Nelson$^{*,\dagger}$, {\em Member, IEEE} and Behrouz Farhang-Boroujeny$^{*}$, {\em Life Senior Member, IEEE}
\thanks{$^*$Electrical and Computer Engineering Department, University of Utah, USA,
$^{\dagger}$Wireless Communications Research Department, Idaho National Laboratory, Idaho Falls, USA}
\thanks{Manuscript received August 27, 2025}}

\maketitle

\begin{abstract}
  The use of non-orthogonal signals has several benefits over orthogonal signals in multi-coded communications. We provide a novel, theoretical study of non-orthogonal signaling to expand the applicability of these schemes. Motivated by a class of multi-carrier spread spectrum systems, this paper presents a thorough symbol error rate analysis of the broad class of multi-code signaling methods when they make use of codes which are not necessarily orthogonal. Our analysis is also extended to the case where the code set includes the negative of each code vector, i.e.,  an extension to biorthogonal signaling. Moreover, it is shown that the symbol error rate results derived in this paper reduce to those available in the literature when the multi-codes are orthogonal or have equal correlation between vectors. Additionally, we show how Monte Carlo integration can be used to evaluate the integrals in the error probability calculation and derive low complexity upper bounds on the error probabilities. {\color{black} We show that by combining these techniques, the error probability can be efficiently computed across the full SNR regime. Finally, we use the upper bound of the error probability to develop some analytical insights about the impacts of non-orthogonality among the code vectors on the symbol error probability.\color{black}}
\end{abstract}

This work has been submitted to the IEEE for possible publication. Copyright may be transferred without notice, after which this version may no longer be accessible.

\section{Introduction}\label{sec:Intro}
\IEEEPARstart{O}{rthogonal} signaling is a well known communication technique, \cite{ProakisJohnG2008Dc, SimonMarvinKenneth1995Dct}. It is widely used to increase the data rate of spread spectrum communication systems through a technique known as multi-coding. A multi-coded system makes use of an $M$-ary symbol alphabet to send information and is closely related to code division multiple access (CDMA) techniques. In CDMA, users are assigned unique codes that allow them to operate simultaneously over a shared channel. In multi-coding, multiple codes are assigned to a single user and can thus be used to increase the data rate of that user. This technique has been well studied and is often referred to as multi-code CDMA \cite{1618816,JinkangZhu1995Papo}.

Prior multi-coding works have used orthogonal signals to define the multi-codes, \cite{HaabDavidB2018FBMS, OhtaTatsuya2013Mfep,1618816}. When the codes are chosen from a set of orthogonal signals, the symbol error probabilities are well-known and documented in textbooks, e.g., \cite{ProakisJohnG2008Dc}. It also has been noted that one may add the negated versions of the orthogonal signals/codes into the multi-code set to double the size of the set, hence, one additional bit is added to each coded transmission. The result is often referred to as biorthogonal signaling and is found to be optimal when the number of signal vectors is twice the dimensionality of the signal space \cite{WeberC.1966Nstt}. Though orthogonal/biorthogonal signaling allows for simple probability of error analysis and may offer optimal performance in some cases, non-orthogonal signal sets can offer other performance advantages that warrant their use in certain applications.  

The constant envelop orthogonal frequency division multiplexing phase modulation, known as OFDM-PM is a non-orthogonal signaling method that has been found to be of great value in certain applications because of its very low peak-to-average power ratio (PAPR), e.g., see \cite{1495013}. The IEEE 802.15.4 wireless standard suggests a non-orthogonal frequency shift keying (FSK) method to improve on the bandwidth efficiency of transmission, \cite{9206104}. Similarly, the authors of \cite{PandeyApurveK.2022EAaO} use non-orthogonal FSK to reduce the bandwidth in a power line communication system. In \cite{HongY.Q.2022NPtf}, non-orthogonal signaling is used to reduce the complexity of an optical communication system.

It has been shown that in an emerging class of filter bank multi-carrier spread spectrum (FBMC-SS) waveforms the use of a set of non-orthogonal multi-codes can provide a number of benefits, with a negligible loss when compared to a set of orthogonal multi-codes. This included a significantly reduced PAPR, a low complexity detector implementation, and perfectly flat power spectral density (PSD) of the synthesized transmit signal, \cite{ofmt_paper}. In an FBMC-SS transceiver, the symbol chips span over a set of subbands of the transmit signal, e.g., see \cite{DarylUCC,LarawayStephenAndrew2020PAoa}. As with the previous works, e.g., \cite{ProakisJohnG2008Dc, SimonMarvinKenneth1995Dct,1618816,JinkangZhu1995Papo,HaabDavidB2018FBMS}, a multi-code signaling scheme can be adopted to improve the spectral efficiency of FBMC-SS.

Prior FBMC-SS designs have utilized filtered-multitone spread spectrum (FMT-SS) where the transmission is over a set of non-overlapping subcarrier bands. In FMT-SS, a method for choosing orthogonal codes with excellent PAPR and low complexity decoding has been provided in \cite{HaabDavidB2018FBMS}. More recently, in \cite{ofmt_paper}, we showed that, by restricting the multi-code chips to the values of $\pm 1$, adjacent subcarrier bands can be overlapped to result in a flat PSD without introducing any inter-carrier interference (ICI). This form of FBMC-SS has been named overlapped FMT-SS (OFMT-SS). OFMT-SS may prove useful across a variety of applications. The particular applications of interest are communications through skywave high-frequency (HF) channels over the bands in the range of  $3$ to $30$~MHz,  the ultra-wideband (UWB) communications, and development of a control channel for spectrum sharing in a cognitive radio setup. In these applications, transmissions may interfere with other users of the same spectrum, hence, measures should be taken to minimize its PSD to the extent possible. Flattening the PSD of OFMT-SS is an important step towards this goal. In \cite{ofmt_paper}, a non-orthogonal multi-coding scheme satisfying the restrictions on multi-code chips (required for keeping its PSD flat) while minimizing the PAPR of the synthesized signal, is proposed.  Here, due to the restrictions on the code elements, the codes are neither orthogonal nor equi-correlated, but have low cross-correlations. Our initial study of the performance of these codes, in terms of their symbol error rates, revealed that they suffer from a minimal loss when compared to their orthogonal counter-parts. This motivated the study presented in this paper.

{\color{black} By gaining a better understanding of the error probabilities of non-orthogonal multi-coding schemes, a system designer is able to understand the cost in error rate performance that he may experience for selecting a non-orthogonal code set. In the case of OFMT-SS, the symbol error rate analysis allows for an informed decision to be made between an orthogonal code with a higher signal PAPR and a scheme with lower PAPR but a non-orthogonal code set. In addition, having theoretical error probabilities allows for a designer to verify the performance of a particular implementation against a baseline.\color{black}}

Beyond a few special cases, theoretical study of the symbol error probabilities of non-orthogonal multi-coding schemes is limited. Reference \cite{1057730} has derived numerically tractable integrals for symbol error probability of multi-coding systems when the cross-correlation of all the codes are equal. These results are applied to find the probability of error for simplex signaling \cite{cd616825-5a73-34ba-a060-a4d5420f325b}. Alternative derivations showing how the results for orthogonal signaling can be applied to simplex signaling are given in \cite{ProakisJohnG2008Dc,BALAKRISHNAN1961485}. For more general cases, others have concentrated on development of lower and upper bounds of symbol error probabilities, \cite{RUGINI2019124,1495013,1091253}. Many of these bounds are based on the work in \cite{1057730} and in some cases have been acknowledged to be loose \cite{6769660}.

In this paper, we study the error probabilities of multi-coding with non-orthogonal codes. Throughout this paper, we refer to these codes as ``quasi-orthogonal'' and ``quasi-biorthogonal.'' These labels are found to be convenient, and representative of the cases of practical interest, though it should be noted that the results are generally applicable to the broad class of non-orthogonal codes.

To study the error probability of non-orthogonal codes, we derive high-dimensional integrals with closed-form integration bounds that provide the exact error probabilities for quasi-orthogonal and quasi-biorthogonal codes. The method used to derive these error probabilities is then extended to the special cases of orthogonal and equi-correlated codes where, as in the prior literature, it is found that a high-dimensional integration is not required for evaluation of the exact error probabilities. To the best of our knowledge, the derivation of error probabilities for quasi-orthogonal and quasi-biorthogonal codes have not previously been presented in the literature. We show how the error probabilities can be evaluated using Monte Carlo integration techniques. We also derive a low-complexity bound based on the pair-wise error probability (PEP) for orthogonal and biorthogonal signaling. This bound is shown to be tight at high signal-to-noise ratio (SNR) values where PEPs drop to very small values. Finally, we present a visualization based on the PEP bound that demonstrates the impact of non-orthogonality on a biorthogonal signaling scheme. Using the methods presented here, the error probabilities for quasi-orthogonal and quasi-biorthogonal signaling schemes can be evaluated in a reasonable computation time across the full range of SNR values.

The rest of this paper is organized as follows.  Section~\ref{sec:quasi-ortho} is devoted to the case of quasi-orthogonal signaling. We show that the probability of a symbol error of such signaling methods can be formulated as an average of a set of multiple integrals. We show that this general result reduces to those reported in the literature for the case of orthogonal signaling by simply plugging in the respective correlation coefficients. For the case of equi-correlated signaling, we show a simple modification to our derivations leads to the final results presented in equation (35) of \cite{1057730}. Section~\ref{sec:quasi-biortho} is devoted to the case of quasi-biorthogonal signaling. Here, we find that the derivation presented earlier for the quasi-orthogonal case can be easily extended to the case of quasi-biorthogonal signaling.  In Section~\ref{sec:mc_evaluation}, we make use of a Monte Carlo integration technique to evaluate the multiple integrals developed in Sections~\ref{sec:quasi-ortho} and \ref{sec:quasi-biortho}, with an affordable computational complexity. Section~\ref{sec:UpperBound} is devoted to derivation of upper bounds for symbol error probabilities of both quasi-orthogonal and quasi-biorthogonal signaling methods. {\color{black}In Section~\ref{sec:discussion}, we use the upper bound to explore the impacts of non-orthogonality on the symbol error probabilities and show that non-orthogonality has very little impact on the error probability for small deviations from the orthogonal signaling case. \color{black}} The numerical results presented in Section~\ref{sec:sims} demonstrate the analysis in the previous sections and show that the upper bounds are very tight at the symbol error rates (SERs) of practical interest ($<10^{-3}$). The concluding remarks of the paper are made in Section~\ref{sec:conclusion}.

\vspace{3mm}

\noindent
{\em Notations:} Matrices, vectors, and scalar quantities are denoted by boldface uppercase, boldface lowercase, and normal letters, respectively. The element in the $i$th row and the $j$th column of a matrix $\bf R$ is denoted as $r_{ij}$, and ${\bf r}_j$ denotes the $j$th column of the same matrix. $\E[\cdot]$ denotes expectation, and ${\bf I}$ refers to the identity matrix. The superscripts $\cdot^{\rm T}$, $\cdot^{\rm H}$, and $\cdot^*$ indicate transpose, conjugate transpose, and conjugate operations, respectively. The Euclidean norm of a vector $\bf v$ is denoted by  $\|\bf v\|$. The notation $\Norm(\bm{\mu}, \bm{\Sigma})$ represents the normal distribution of a random vector with mean $\bm{\mu}$ and covariance matrix  $\bm{\Sigma}$.

\section{Quasi-Orthogonal Signaling}\label{sec:quasi-ortho}

\subsection{System Model}\label{sec:system_model}
We consider a communication system where the transmit information symbols are chosen from a set of real-valued length $L$ alphabet vectors $\left\{ \s _0, \s _1, \ldots, \s _{M-1}\right\}$ which may not be orthogonal, but are linearly independent.
Moreover, for $i=0,\ldots,M-1$, $\|\s_i\| = 1$. In addition, all symbol vectors are equally likely to be transmitted.

We consider an additive white Gaussian noise (AWGN) channel, where, when $\s_i$ has been transmitted, the received signal vector is given as
\begin{equation} \label{eq:rx_vector}
  \x = \sqrt{\En}\s _i + {\bf n},
\end{equation}
where $\En$ is the received symbol energy, and ${\bf n}$ is a channel noise vector with distribution ${\bf n} \sim \Norm\left(\bm 0, \frac{N_0}{2}\textbf{I}\right)$.

To detect the transmitted information, we make use of  the maximum likelihood (ML) decision rule, \cite{ProakisJohnG2008Dc}, where the receiver chooses
\begin{equation}\label{eq:MLrule}
  \hat\s = \arg\max_{\s_i} \; \s_i^\T \x.
\end{equation}

\subsection{Symbol Error Probability Analysis} \label{sec:general_symbol_error}

For the $M$-ary signaling discussed above, the probability of a symbol error may be given as
\begin{equation} \label{eq:P_e}
  P_e  = 1 - \frac{1}{M}\sum_{i=0}^{M-1}P_{c|\s_i}
\end{equation}
where $P_{c|\s_i} $ refers to the probability of correct detection when the symbol vector  $\s_i$ has been transmitted.  We note that because of different correlations among different symbol vectors, $P_{c|\s_i}$ should be calculated separately for all choices of $0\le i\le M-1$.

Next, we note that
\begin{equation}\label{eq:Decision_Integral1}
  P_{c|\s_i} =  \int\cdots\int_{D_i} f_{X}(\x) d\x,
\end{equation}
where $D_i$ indicates the  decision region that corresponds to the ML decision rule \eqref{eq:MLrule}, and $f_{X}(\x)$ is the probability density function (PDF) of $\x$.

Note that in \eqref{eq:Decision_Integral1}, the integral has the same dimension as the size of the signal vector $\x$, i.e., it is an $M$-fold integral. Computation of such integrals, even numerically, in general, may not be a straightforward task. In the sequel, we show how the integrals in \eqref{eq:Decision_Integral1} may be rearranged to allow numerical evaluation of $P_{c|\s_i}$, for all practical choices of $M$, that may be as large as a few hundreds.

We order the variables of integration so that the limits of the outer integrals do not depend on the variables of integration for the inner integrals. To this end, for a given choice $\s_i$, we organize the signal vectors $\left\{\s_0, \s_1, \ldots, \s_{M-1}\right\}$ into a matrix
\begin{equation}\label{eq:S}
  \SM_i = \begin{bmatrix}
    \s_i & \s_{i + 1} & \cdots & \s_{M-1} & \s_0 & \s_1 & \cdots \s_{i-1}
  \end{bmatrix}
\end{equation}
of dimension $L\times M$. Let $\SM_i =\Q\R$ be the QR factorization of $\SM_i$, where $\Q$ is an $L \times M$ matrix whose columns are an orthonormal basis set and $\R$ is an $M \times M$ upper triangular matrix.

Before deriving the integration bounds, we make a few observations about $\Q$ and $\R$. First, we note that the QR factorization is not unique and can always be chosen so that all diagonal elements of $\R$ are non-negative.  Moreover, for this choice of the QR factorization, one may choose to set $r_{00} = 1$. In that case, the first column of $\Q$ will be equal to $\s_i$. Also, one may note that the orthonormality of the columns of $\Q$ implies that the columns of $\R$ are a set of unit length vectors, i.e., $\|\rv_i\|=1$, for $0\le i\le M-1$.

Next, making use of \eqref{eq:rx_vector}, we define the length $M$ vector
\begin{align} \label{eq:transformed_rx_signal}
  \y & = \Q^\T \x                   \nonumber \\
     & = \sqrt{\En} \rv_0 + {\bf n}'
\end{align}
where ${\bf n}'=\Q^\T{\bf n}$. Recalling that the columns of $\Q$ are a set of  orthonormal vectors, one finds that
\begin{equation} \label{eq:transformed_pdf}
  \y \sim \Norm \left(\sqrt{\En} \rv_0, \frac{N_0}{2}\textbf{I} \right).
\end{equation}
Moreover, \eqref{eq:Decision_Integral1} may be rewritten as
\begin{equation} \label{eq:P_c_unsimplified}
  P_{c|\s_i} = \int\cdots\int_{D'_i} f_Y(\y) d \y
\end{equation}
where $D'_i$ is the decision region for correct detection of $\rv_0$, and $f_Y(\y)$ is the PDF of $\y$, given in \eqref{eq:transformed_pdf}.

Recalling \eqref{eq:S}, for any $j\ne i$, $\s_j$ is transformed to
\begin{equation}
  \rv_{j'} = \Q^\T \s_j
\end{equation}
where
\begin{equation}
  j' = \begin{cases}
    j - i ,  & j > i  \\
    M-i + j, & j < i.
  \end{cases}
\end{equation}
Taking note that the vectors $\rv_0$ and $\rv_{j'}$ have unit length, the decision boundary between $\rv_0$ and $\rv_{j'}$ will be the hyperplane, \cite{BALAKRISHNAN1961485},
\begin{equation}
  (\rv_0 - \rv_{j'})^\T \y=0.
\end{equation}
Hence, a correct decision is made when
\begin{equation}\label{eq:upper_bound}
  (\rv_0 - \rv_{j'})^\T \y > 0.
\end{equation}

Next, taking note that  $\R$ is upper triangular and $r_{00}=1$,  \eqref{eq:upper_bound}  can be written as $y_0 - \sum_{i=0}^{j'} r_{k,j'}y_i > 0$, or, equivalently,
\begin{equation} \label{eq:ub_w_func}
  y_{j'}  <  u_{j'}(y_0, y_1, \ldots, y_{j'-1})
\end{equation}
where
\begin{equation} \label{eq:ub_func}
  u_{j'}(y_0, y_1, \ldots, y_{j'-1}):=\frac{y_0 - \sum_{k=0}^{j'-1} r_{k,j'}y_k}{r_{j',j'}}.
\end{equation}
This is the upper limit of the $(j'+1)^{\text{th}}$ inner integral in \eqref{eq:P_c_unsimplified}. Using these bounds, \eqref{eq:P_c_unsimplified} can be written
\begin{equation}\label{eq:Px|si}
  P_{c|\s_i} = \int_{-\infty}^\infty \int_{-\infty}^{u_1(y_0)} \cdots \int_{-\infty}^{u_{M-1}(y_0, y_1, \ldots, y_{M-2})} f_Y(\y) d\y.
\end{equation}

To convert \eqref{eq:Px|si} to a simpler form, we recall the distribution \eqref{eq:transformed_pdf} and accordingly introduce the following change of variables
\begin{equation} \label{eq:non_o_sn_var0}
  v_0 = \frac{y_0 - \sqrt{\En}}{\sqrt{\frac{N_0}{2}}}
\end{equation}
and
\begin{equation} \label{eq:non_o_sn_vari}
  v_i = \frac{y_i}{\sqrt{\frac{N_0}{2}}},~~\mbox{for $i = 1, 2, \ldots, M - 1$,}
\end{equation}
to rearrange \eqref{eq:Px|si} as in \eqref{eq:p_c_integral} at the top of the next page.
\begin{table*}
  \normalsize  \begin{equation} \label{eq:p_c_integral}
    P_{c|\s_i} =  \int_{-\infty}^\infty f(v_0) \int_{-\infty}^{u'_1(v_0)} f(v_1) \cdots        \int_{-\infty}^{u'_{M-1}(v_0, v_1, \ldots, v_{M-2})} f(v_{M-1}) dv_{M-1}\cdots dv_1dv_0,
  \end{equation}
  where
  \normalsize \begin{equation} \label{eq:standard_normal_ub}
    u'_j(v_0, v_1, \ldots, v_{j-1}) = \frac{1}{r_{j,j}}\left[(1 - r_{0,j})\left(v_0 + \sqrt{\frac{2 \En}{N_0}}\right) - \sum_{k=1}^{j-1}r_{k,j}v_k \right]
  \end{equation}
  \hrulefill
\end{table*}
In \eqref{eq:p_c_integral}, $f(x)$ is the standard normal function defined as

\begin{equation}
  f(x)=\frac{1}{\sqrt{2 \pi}} e^{-\frac{1}{2}x^2}.
\end{equation}

\subsection{Extensions to the Known Cases}
In the current literature one will find equations for the SERs of multi-codes for a couple of specific cases. Here, we show how the above results of the general case may be simplified to reduce to those in the literature.

\subsubsection{Orthogonal Case}\label{sec:orth case}
When the code set $\left\{ \s _0, \s _1, \ldots, \s _{M-1}\right\}$ consists of a set of orthogonal vectors, the QR decomposition of $\SM$ has the trivial solution $\SM\I$. That is, $\R$ is the identity matrix $\I$, hence, $r_{j,j}=1$ and for $j<k$, $r_{j,k}=0$. For this particular case, \eqref{eq:standard_normal_ub} reduces to
\begin{equation}\label{eq:standard_normal_ub_orth}
  u'_j(v_0, v_1, \ldots, v_{j-1})=v_0 + \sqrt{\frac{2 \En}{N_0}},
\end{equation}
whose application in \eqref{eq:p_c_integral} leads to the well known result \cite[pg. 205]{ProakisJohnG2008Dc}
\begin{equation} \label{eq:ortho_Pc}
  P_{c|\s_i} = \int_{-\infty}^\infty f(v_0) \left( \int_{-\infty}^{v_0 + \sqrt{\frac{2\En}{N_0}}} f(v_1) dv_1 \right)^{M-1} dv_0.
\end{equation}

\subsubsection{Equi-Correlated Case}\label{sec:equicorr}
The equi-correlated case that was first studied in \cite{1057730} considers the case where
\begin{equation}\label{eqn:SS}
  \SM^\T \SM = \W = \begin{bmatrix}
    1      & \eta   & \cdots & \eta \\
    \eta   & 1      & \cdots & \eta \\
    \vdots & \vdots & \ddots & \eta \\
    \eta   & \eta   & \cdots & 1
  \end{bmatrix}.
\end{equation}
We take note that the matrix $\SM$ is of size $L\times M$ and its columns are the alphabet vectors $\s _0$, $\s _1$, $\ldots$, and $\s _{M-1}$. Accordingly, $\eta$ is the correlation between any pair of these alphabet vectors and $\W$ is a positive semi-definite matrix of size $M\times M$. In simplex signaling $\eta=-\frac{1}{M-1}$ and $L=M$.

Consider the eigen-decomposition
\begin{equation}
  \W = \V \bm{\Lambda}\V^\T,
\end{equation}
where the columns of $\V$ are the eigenvectors of $\W$ and $\bm{\Lambda}$ is a diagonal matrix with eigenvalues of $\W$ at its diagonal.
Taking note that
\begin{equation}\label{eqn:W1}
  \W = \eta \bm{1} \bm{1}^\T + (1 - \eta)\Id,
\end{equation}
it is straightforward to show that the eigenvalues of $\W$ are $\lambda_0=1+(M-1)\eta$, and for $1\le i\le M-1$, $\lambda_i=1-\eta$. We also take note that $\bm{\Lambda}$ can be written as
\begin{equation}
  \bm{\Lambda} = \eta M \B + (1 - \eta)\Id,
\end{equation}
where $\B=\mbox{diag}\left([1,0,\cdots,0]\right)$.

Next, we recall that a square root of $\W$ can be found as
\begin{align}
  \textbf{U} = \V \bm{\Lambda}^{\frac{1}{2}} \V^\T.
\end{align}
Like the matrix $\bm\Lambda$, the matrix $\bm{\Lambda}^{\frac{1}{2}}$ can be also split into two matrices as
\begin{equation}
  \bm{\Lambda}^{\frac{1}{2}} = a \B + b \Id
\end{equation}
where
\begin{equation}
  a = \sqrt{(M - 1)\eta + 1} - \sqrt{1 - \eta}
\end{equation}
and
\begin{equation}
  b = \sqrt{1 - \eta}.
\end{equation}
Making use of this observation, one finds that
\begin{equation} \label{eq:equi_corr_mat}
  \textbf{U} =  \begin{bmatrix}
    d      & c      & \cdots & c \\
    c      & d      & \cdots & c \\
    \vdots & \vdots & \ddots & c \\
    c      & c      & \cdots & d
  \end{bmatrix},
\end{equation}
where
\begin{equation}
  c = \frac{1}{M}\left( \sqrt{1 + \eta(M-1)}-\sqrt{1 - \eta} \right)
\end{equation}
and
\begin{equation}
  d  = \sqrt{1 - \eta} + c.
\end{equation}

Considering the above results, one may conclude that the matrix $\SM$ that carries the symbol alphabets as its columns has the general form
\begin{equation}\label{eq:S matrix}
  \SM=\Q\U
\end{equation}
where $\Q$ is a matrix of size $L\times M$ with a set of unit length orthogonal vectors as its columns. Here, given the simple and special form of the matrix $\U$, we make use of the factorization \eqref{eq:S matrix} instead of the QR factorization.

We also take note that given the symmetry of the matrix $\U$, the symbol error probabilities are independent of the choice of the transmitted symbol. Hence, without any loss of generality, we consider the case where $\s_0$ has been transmitted. Thus, using \eqref{eq:rx_vector}, with $\s_i$ replaced by $\s_0$, one finds that
\begin{align} \label{eq:transformed_rx_signal2}
  \y & = \Q^\T \x                   \nonumber \\
     & = \sqrt{\En} \uv_0 + {\bf n}'
\end{align}
where  ${\bf n}'=\Q^{\rm T}{\bf n}$.

Here, the decision boundary for making a correct decision is found to be
\begin{equation} \label{eq:equi-corr_bound}
  (\uv_0 - \uv_i)^\T \y > 0.
\end{equation}
Substituting the relevant columns of $\U$ in \eqref{eq:equi-corr_bound}, and taking note that $d-c=\sqrt{1-\eta}$ is a positive number, \eqref{eq:equi-corr_bound} reduces to
\begin{equation}
  y_i < y_0,
\end{equation}
for $1 \le i \le M - 1$. This leads to the following equation for the probability of a correct decision
\begin{equation} \label{eq:equi-corr_pc}
  P_c = \int_{-\infty}^{\infty} f_{Y_0}(y_0) \left(\int_{-\infty}^{y_0} f_{Y_1}(y_1)dy_1\right)^{M - 1} dy_0.
\end{equation}
\begin{table*}
  \normalsize \begin{equation} \label{eq:p_c_integral_biortho}
    P_{c|\s_i} =  \int_{-\sqrt{\frac{2 \En}{N_0}}}^\infty f(v_0) \int_{l'_1(v_0)}^{u'_1(v_0)} f(v_1) \cdots        \int_{l'_{M-1}(v_0, v_1, \ldots, v_{M-2})}^{u'_{M-1}(v_0, v_1, \ldots, v_{M-2})} f(v_{M-1}) dv_{M-1}\cdots dv_1 dv_0,
  \end{equation}
  \normalsize where $u'_j(v_0, v_1, \ldots, v_{j-1})$ is given in \eqref{eq:standard_normal_ub} and
  \normalsize \begin{equation} \label{eq:standard_normal_lb}
    l'_j(v_0, v_1, \ldots, v_{j-1}) = \frac{1}{r_{j,j}}\left[-(1 + r_{0,j})\left(v_0 + \sqrt{\frac{2 \En}{N_0}}\right) - \sum_{k=1}^{j-1}r_{k,j}v_k \right].
  \end{equation}
  \hrulefill
\end{table*}
Recognizing that
\begin{equation}
  \y \sim \Norm \left(\sqrt{\En} \uv_0, \frac{N_0}{2}\Id \right)
\end{equation}
and applying the conversions to standard normal following \eqref{eq:non_o_sn_var0} and \eqref{eq:non_o_sn_vari} in \eqref{eq:equi-corr_pc}, we get
\begin{equation} \label{eq:equi_corr_pc}
  P_c = \int_{-\infty}^{\infty} f(v_0) \left(\int_{-\infty}^{v_0 + \sqrt{\frac{2\En(1-\eta)}{N_0}}} f(v_1)dv_1\right)^{M - 1} dv_0.
\end{equation}
This result is similar to that of the orthogonal case with the received signal energy scaled by a factor equal to $1-\eta$. This result is in perfect agreement with the probability of correct symbol detection reported in  \cite{1057730}; see equation (35) and the following paragraph in this paper.

{\color{black}Though \eqref{eq:equi_corr_pc} has also been derived in \cite{1057730}, the derivation here is considerably different, hence, merits a presentation in the context of this paper. In \cite{1057730}, the author begins with the noise terms after correlating the received signal with the bank of multi-codes. It is then shown that these noise terms are correlated, and their covariance matrix has the same form  as the matrix $\W$. Subsequently, making use of the special form of the cofactors of $\W$ and  some clever steps, a result similar to \eqref{eq:equi_corr_pc} is arrived.
\color{black}}

\section{Quasi-Biorthogonal Signaling}\label{sec:quasi-biortho}
\subsection{System Model}
Here, we extend the above results to the case where the alphabet vectors are extended to the following set
\begin{align}
  \s _0     & = -\s _M      \nonumber \\
  \s _1     & = -\s _{M+1}  \nonumber \\
            & \vdots        \nonumber \\
  \s _{M-1} & = -\s _{2M-1}.
\end{align}
As before, the vectors $\s_0$, $\s_1$, $\cdots$, $\s_{M-1}$ may not be orthogonal,  but are linearly independent. We also assume that the detector uses the maximum likelihood (ML) decision rule, \cite{ProakisJohnG2008Dc}, where the receiver first computes
\begin{equation}
  z_i = \s_i^\T \x \text{ for } 0 \le i <M,
\end{equation}
and subsequently obtains the estimate of the transmitted symbol as
\begin{equation}
  \hat{\s } = \sign({z_k}) \s_k,~~\mbox{where }k = \arg\max_{0 \le i < M}|z_i|.
\end{equation}

\subsection{Symbol Error Probability Analysis}
The probability of a decision error can be found using \eqref{eq:P_e}, but due to the symmetry of the signal set, the probability of error is the same for both $\s_i$ and $-\s_i$. Hence, \eqref{eq:P_e} is applicable to the case here as well. Moreover, as in the case of orthogonal signaling, here also $P_{c|\s_i}$ can be found by finding the decision region and integrating the PDF over that region. The decision boundaries are obtained by following the same line of derivations as in Section~\ref{sec:general_symbol_error}. The difference here is that in addition to the upper bounds in the integrals, there are also a set of lower bounds. While the upper bounds are those obtained using \eqref{eq:upper_bound}, the lower bounds are obtained using the inequality
\begin{equation}\label{eq:lower_bound}
  (\rv_0 + \rv_{j'})^\T \y > 0.
\end{equation}
Following the same line of derivations to those in Section~\ref{sec:general_symbol_error}, one will find that \eqref{eq:lower_bound} leads to
\begin{equation}
  y_{j'}  >  l_{j'}(y_0, y_1, \ldots, y_{j'-1}),
\end{equation}
where
\begin{equation}
  l_{j'}(y_0, y_1, \ldots, y_{j'-1}):= \frac{-y_0 - \sum_{k=0}^{j'-1} r_{k,j'}y_k}{r_{j',j'}}.
\end{equation}
In addition, here, the detector should distinguish between $\rv_0$ and $-\rv_0$. The decision boundary between $\rv_0$ and $-\rv_0$ is the hyperplane orthogonal to $\rv_0$ and passing through the origin. This can be written as
\begin{equation}
  y_0>0.
\end{equation}
This specifies the limits of integration in the outer integral.

Considering the above points, one will find that
\begin{equation}\label{eq:Px|si_biortho}
  P_{c|\s_i} = \int_{0}^\infty \int_{l_1(y_0)}^{u_1(y_0)} \cdots \int_{l_{M-1}(y_0, y_1, \ldots, y_{M-2})}^{u_{M-1}(y_0, y_1, \ldots, y_{M-2})} f_Y(\y) d\y.
\end{equation}
Making use of \eqref{eq:non_o_sn_var0} and \eqref{eq:non_o_sn_vari}, this result can be rewritten in terms of the standard normal PDFs. The result is presented in  \eqref{eq:p_c_integral_biortho} at the top of the current page.

For the case where the symbol vectors $\s_0$ through $\s_{M-1}$ are an orthogonal set,  $\R$ will be an identity matrix (see the discussion in Section~\ref{sec:orth case}) and as a result \eqref{eq:p_c_integral_biortho} reduces to
\begin{equation} \label{eq:biortho_Pc}
  \begin{aligned}
    P^'_{c|\s_i} = & \int_{-\sqrt{\frac{2 \En}{N_0}}}^\infty f(v_0) \left( \int_{-v_0 - \sqrt{\frac{2\En}{N_0}}}^{v_0 + \sqrt{\frac{2\En}{N_0}}} f(v_1) dv_1 \right)^{M-1} dv_0.
  \end{aligned}
\end{equation}
This is in perfect agreement with those reported in the literature; e.g., see \cite[pg. 208]{ProakisJohnG2008Dc}.

\section{Numerical Calculation $P_{e}$} \label{sec:mc_evaluation}
To determine $P_{e}$, we must evaluate the integral given in \eqref{eq:p_c_integral}, in the case of quasi-orthogonal signaling, or \eqref{eq:p_c_integral_biortho}, in the case of quasi-biorthogonal signaling and apply \eqref{eq:P_e}. The dimensionality of the integrals is $M$, with no apparent method for simplification. To evaluate the integrals, we turn to Monte Carlo integration techniques. This approach has been applied across a wide variety of disciplines and has proven effective in efficiently evaluating high dimensional integrals \cite{KalosMalvinH2008MCm}. Here, we show how the integral in \eqref{eq:p_c_integral} or \eqref{eq:p_c_integral_biortho} can be formulated and solved to a desired precision using Monte Carlo integration techniques.

{\color{black} To evaluate an integral using Monte Carlo techniques, we formulate an expression for the integral as the expected value of a function $g(X)$ of a random variable, and then design an estimator of that function to approximate the value of the integral. Using this method, the Chebychev inequality can be used to determine the number of samples required for the estimator to achieve an acceptable level of performance. In this section, we apply this theory to our particular integration problem. \color{black}} If the estimator is consistent, its value will approach the expected value as the number of samples increases.

In our case, we define the functions
\begin{equation}\label{eq:gi(V)}
  g_i(\textbf{v}) = \begin{cases}
    1 & {\textbf{v}} \in D_i \\
    0 & \text{otherwise,}
  \end{cases} \quad \text{for }0 \le i \le M - 1
\end{equation}
where $D_i$, as defined before, refers to the domain of the multidimensional integral in \eqref{eq:p_c_integral} or \eqref{eq:p_c_integral_biortho}. Accordingly, the desired estimator for each $P_{c|\s_i}$ is formulated as
\begin{align} \label{eq:mc_estimator_i}
  P_{c|\s_i} & = \int_{-\infty}^{\infty} \cdots \int_{-\infty}^{\infty} g_i({\bf v}) f_{{\bf V}}( {\bf v}) d {\bf v} \nonumber \\
             & \approx \frac{1}{K} \sum_{k = 0}^{K - 1} g_i({\bf V}_k) \nonumber                                               \\
             & := G_i,
\end{align}
where $f_{\bf V}(\bf v)$ is the multivariate Gaussian PDF $\Norm(\bm 0, \frac{N_0}{2}\textbf{I})$ and $G_i$ indicates the estimator result for $P_{c|\s_i}$. Because the desired result is $P_e$, we form an estimator of $P_c$ as
\begin{equation} \label{eq:mc_estimator_full}
  G = \frac{1}{M}\sum_{i=0}^{M-1}G_i.
\end{equation}
To find $G$, we evaluate $g_i(\Vb)$ by drawing $K$ independent and identically distributed (IID) samples from $f_{\bf V}(\bf v)$, denoted by ${\bf v}_k$,  and insert the results in  \eqref{eq:mc_estimator_i}. The function $g_i(\bf v)$ is evaluated following \eqref{eq:gi(V)}, by checking if  $\bf v$ is within the bounds defined by $D_i$. This process is repeated for each $\s_i$ and the estimate of $P_c$ is computed using \eqref{eq:mc_estimator_full}. Hence, the total number of samples of $\bf v$ used to obtain an estimate of $G$ is $KM$.

{\color{black}To find the number of samples $K$ required per estimator $G_i$ for a sufficient accuracy, we follow the standard procedure of applying the Chebychev inequality, \cite[pg. 23]{KalosMalvinH2008MCm}.\color{black}} For the case here, the Chebychev inequality states that
\begin{equation} \label{eq:cheb_eq}
  \Pr\left\{|G - \E[G]| \ge \left[ \frac{\var{\left[G\right]}}{\delta} \right]^{\frac{1}{2}} \right\} \le \delta,
\end{equation}
where $\delta$ is a (small) constant and $\var[\cdot]$ denotes variance of. To proceed, we first note that \eqref{eq:mc_estimator_i} and \eqref{eq:mc_estimator_full} imply
\begin{align}\label{eqn:varG}
  \var[G] & = \frac{1}{KM^2}\sum_{i=0}^{M-1}\var[g_i(\Vb)] \nonumber                           \\
          & = \frac{1}{KM}\stackunder{\text{avg}}{\scriptsize $i$}\left(\var[g_i(\Vb)]\right).
\end{align}
Moreover, since $g_i(\Vb)$ is a binary random variable,  it has a Bernoulli distribution with parameter $p=P_{c|\s_i}=1-P_{e|\s_i}$, hence, $\var[g_i(\Vb)] = P_{e|\s_i}(1-P_{e|\s_i})$.

Taking note that our interest is mostly centered around the points where the probability $P_{e|\s_i}$ is small, say $P_{e|\s_i}<0.01$, one may argue $P_{e|\s_i}(1-P_{e|\s_i})\approx P_{e|\s_i}$ and, hence,
\begin{equation}\label{eqn:avg gi(V)1}
  \stackunder{\text{avg}}{\scriptsize $i$}\left(\var[g_i(\Vb)]\right) \approx P_e.
\end{equation}

Another point that should be taken note of is the fact that $\E[G]=1-P_e$ and $G=1-\hat P_e$, where $\hat P_e$ is an estimate of $P_e$. These imply
\begin{equation}\label{eqn:G=Pe}
  |G - \E[G]|=|P_e-\hat P_e|.
\end{equation}
Making use of \eqref{eqn:varG}, \eqref{eqn:avg gi(V)1} and \eqref{eqn:G=Pe},  \eqref{eq:cheb_eq} may be rearranged as
\begin{align} \label{eq:cheb_bnd}
  \Pr\left\{|P_e - \hat P_e| \ge \left[ \frac{P_e}{KM\delta} \right]^{\frac{1}{2}} \right\} \le \delta.
\end{align}
Considering \eqref{eq:cheb_bnd}, one may argue for $\hat P_e$ to be a close estimate of $P_e$ (equivalently, $G$ be a good approximation to $P_{c}=\E[G]$), one may let
\begin{equation}\label{eq:K1}
  \left[\frac{P_e}{KM\delta} \right]^{\frac{1}{2}}=\epsilon P_e,
\end{equation}
where $\epsilon$ is a small constant. Solving \eqref{eq:K1} for $KM$, i.e., the number of samples required to obtain a good estimate of $P_e$, we get
\begin{equation}\label{eq:KM2}
  KM=\frac{1}{\delta\epsilon^2 P_e}.
\end{equation}

In \eqref{eq:KM2}, we  are facing a problem where $KM$, on the left-hand side, depends on the quantity $P_e$, on the right-hand side, which we are seeking to estimate. This dilemma may be solved by replacing $P_e$ by the known probability of symbol error, $p_e$, from an equivalent orthogonal multi-code. That is, $KM$ is obtained using
\begin{equation}\label{eq:KM3}
  KM=\frac{1}{\delta\epsilon^2 p_e},
\end{equation}
instead of \eqref{eq:KM2}. Since, $p_e<P_e$, \eqref{eq:KM3} may be thought as an over-estimate of $KM$.

{\color{black} As is typical with Monte Carlo integration techniques,  $KM$ is independent of the number of integrals in  \eqref{eq:p_c_integral} or \eqref{eq:p_c_integral_biortho}. \color{black}} To get a feel of how large $KM$ should be, we note that the choices of $\delta=\epsilon=0.01$ are reasonable, and with these choices, \eqref{eq:KM3} reduces to
\begin{equation}\label{eq:KM4}
  KM=\frac{10^6}{p_e}.
\end{equation}
Then, when $p_e=0.01$, $KM=10^8$, and this value increases by an order of magnitude for each decrement of $p_e$ by an order of magnitude. This allows measuring $P_e$  values as low as $10^{-6}$ in a reasonable amount of time, even for the cases where $M$ grows to a few hundreds. {\color{black} Application of the theory in Section~\ref{sec:quasi-ortho} and Section~\ref{sec:quasi-biortho}, combined with this Monte Carlo method allow for a rigorous evaluation of the error probabilities at low SNR in a reasonable computation time. At high SNR, evaluation may not be feasible. To address this and provide accurate symbol error probabilities across the full range of SNR, we explore an upper bound that is tight at high SNR. Combining this bound with the Monte Carlo integration technique, the error probability can be estimated accurately across the full range of SNR. \color{black}}

\section{Development of an Upper Bound}\label{sec:UpperBound}
Although the probability of a symbol error can be found using the methods discussed in Sections~\ref{sec:quasi-ortho} and \ref{sec:quasi-biortho} and numerically evaluated as discussed in Section~\ref{sec:mc_evaluation}, it may be still desirable if one could avoid evaluation of the Monte Carlo integrals. In this section, we derive an upper bound that becomes exceedingly tight over the SNR range of practical interest.  This derivation builds on an alternative derivation (different from those developed in Section~\ref{sec:quasi-ortho} and Section~\ref{sec:quasi-biortho}) of the exact probability of error for the quasi-biorthogonal signaling when $M=2$. The alternative derivation presented for this spacial case is also instructive in visualizing the robustness (a minimum performance loss) of quasi-orthogonal/biorthogonal signaling methods when compared to their  orthogonal/biorthogonal counterparts.

\subsection{Quasi-Biorthogonal Signaling for the Case $M=2$} \label{sec:M4_special_case}

\begin{figure}
  \centering
  \input{biortho_vectors.pgf}
  \caption{The signal vectors, decision boundaries, and an example of the received signal vector when  the signal set is $\left\{ \pm \s_0, \pm\s_1 \right\}$. In (a), these vectors are shown in the original space, and (b) shows the vectors in the coordinate system of ${\bf d}_0$ and ${\bf d}_1$. In both cases the decision region for $+\s_0$ is shaded.} 
  \label{fig:biortho_vectors}
\end{figure}

For this special case, the set of signaling alphabets are the vectors $\left\{\pm\s_0, \pm \s_1 \right\}$ of length $L \ge 2$. Without any loss of generality, consider the case where $+\s_0$ has been transmitted. Then, the received signal is
\begin{equation}\label{eq:x_case M=2}
  \x = \sqrt{\En}\s_0 + {\bf n}.
\end{equation}
In the presentation that follows, we concentrate on the signal vectors within the plane spanned by the vectors $\s_0$ and $\s_1$. We note that, for $L>2$, the noise vector ${\bf n}$ may have components both in the plane spanned by the vectors $\s_0$ and $\s_1$ and perpendicular to it. We also take note that the latter component of ${\bf n}$ has no impact on the symbol decision in the case of interest here and, thus, may be removed from the right-hand side of \eqref{eq:x_case M=2}. Considering this, we assume $\bf n$ in   \eqref{eq:x_case M=2} refers to the component of the noise within the plane spanned by the vectors $\s_0$ and $\s_1$.

The decision boundaries around $\s_0$ are in the direction of the lines $\dv_0^{\T}\x = 0$ and ${\bf d}_1^{\T}\x = 0$ where
\begin{equation} \label{eq:decision_bounds1}
  {\bf d}_0 = \frac{\s_0 - \s_1}{\Vert \s_0 - \s_1\Vert}
\end{equation}
and
\begin{equation} \label{eq:decision_bounds2}
  {\bf d}_1 =\frac{ \s_0 + \s_1}{\Vert \s_0 + \s_1\Vert}.
\end{equation}
These decision boundaries and a choice of the received signal vector $\x$ are presented in Fig.~\ref{fig:biortho_vectors}(a) along with the shaded decision region to correctly select $+\s_0$. Here, we see that although $\s_0$ and $\s_1$ are not orthogonal, the decision boundaries are orthogonal and span the same space as the one spanned by $\s_0$ and $\s_1$. The orthogonality of ${\bf d}_0$ and ${\bf d}_1$ can be verified by observing that ${\bf d}_0^{\T} {\bf d}_1 = 0$.

To evaluate the probability of a symbol error, we start by defining the matrix
\begin{equation}
  \text{\bf D} = \begin{bmatrix}
    {\bf d}_0 & {\bf d}_1
  \end{bmatrix},
\end{equation}
and use the columns of this matrix as an orthogonal basis set to evaluate the probability of a symbol error. This is done by finding $\x$ in the coordinates defined by the basis set ${\bf d}_0$ and ${\bf d}_1$. This is obtained as
\begin{equation}
  \begin{aligned}
    \y & = \text{\bf D}^\T \x                         \\
       & = \sqrt{\En} \text{\bf D}^\T \s_0 + {\bf n}'
  \end{aligned}
\end{equation}
where $\y \sim \Norm(\sqrt{\En}\text{\bf D}^\T \s_0, \frac{N_0}{2}\text{I})$. This operation can be seen in Fig.~\ref{fig:biortho_vectors}(b), where we see that the decision boundaries form the basis axes of the transformed signal space. Accordingly, the probability of a correct symbol decision can be calculated as
\begin{equation} \label{eq:P_c_not_SN}
  P_c = \int_{0}^{\infty} \int_{0}^{\infty} f(\y) d\y.
\end{equation}
To evaluate this probability, we let
\begin{equation}
  {\bf z} = \frac{\y - \sqrt{\En} \textbf{D}^\T \s_0}{\sqrt{\frac{N_0}{2}}}
\end{equation}
and note that the elements  $z_0$ and $z_1$ of $\bf z$ are a pair of independent, zero-mean, and unit variance random variables.  Making use of this result in \eqref{eq:P_c_not_SN}, one will find that
\begin{equation}\label{eq:Pc1}
  P_c = \left(\int_{-\sqrt{\frac{2\En \rho_0^2}{N_0}}}^{\infty} f(z_0) dz_0\right)\left(  \int_{-\sqrt{\frac{2\En\rho_1^2}{N_0}}}^{\infty}  f(z_1)dz_1\right),
\end{equation}
where $\rho_k = {\bf d}_k^\T \s_0$ for $k = 0, 1$, and $f(\cdot)$, as defined earlier, is the standard normal function.
Making use of the $Q$ function, \cite[pg. 21]{KayStevenM.1998Foss}, \eqref{eq:Pc1} reduces to
\begin{equation} \label{eq:P_c_SN}
  P_c = \prod_{k=0}^{1}\left[1 - Q\left(\sqrt{\frac{2\En\rho_k^2}{N_0}}\right)\right],
\end{equation}
Hence, the probability of a symbol error is given by
\begin{equation} \label{eq:P_e_M4}
  \begin{aligned}
    P_e & = 1 - P_c                                                                            \\
        & = 1 - \prod_{k=0}^{1}\left[1 - Q\left(\sqrt{\frac{2\En\rho_k^2}{N_0}}\right)\right].
  \end{aligned}
\end{equation}

At this point it is instructive to look at the symbol constellation points resulting from the code vectors $\s_0$ and $\s_1$ in the Cartesian plane introduced by the basis set ${\bf d}_0$ and ${\bf d}_1$. This is visualized in Fig.~\ref{fig:non_ortho_QPSK}, where the quasi-biorthogonal signal vectors are chosen with $\s_0^\T\s_1=0.25$. When $\s_0$ and $\s_1$ are orthogonal, the constellation points resemble those of a quadrature phase shift keying (QPSK) modulation. Any non-orthogonality of $\s_0$ and $\s_1$ results in a rotation of QPSK points along the unit circle towards either the horizontal or the vertical axis.

The rotation of the QPSK points can be calculated by making use of the equation
\begin{equation}\label{eq:theta}
  \theta = \cos^{-1}\left(\frac{1}{\sqrt{2}}\left(\textbf{d}_0 +\textbf{d}_1\right)^\T \s_0 \right).
\end{equation}
This is the angle between the unit length vectors $\s_0$ and the bisect of basis vectors ${\bf d}_0$ and ${\bf d}_1$. In the case presented in  Fig.~\ref{fig:non_ortho_QPSK}, where  $\s_0^\T\s_1=0.25$, one will find that $\theta=7.2^{\circ}$. As one may see from Fig.~\ref{fig:non_ortho_QPSK}, this translates to a small performance loss for such a relatively large loss of orthogonality. This is in line with the analysis presented in the next section and the simulations in Section~\ref{sec:sims}.

\begin{figure}
  \centering
  \input{biortho_constellation.pgf}
  \caption{Constellation in terms of the basis formed by the decision boundaries $\left[{\bf d}_0 \; {\bf d}_1\right]$ when the signal vectors are orthogonal and not orthogonal.}
  \label{fig:non_ortho_QPSK}
\end{figure}

\subsection{Upper Bound for Quasi-Biorthogonal Signaling} \label{sec:biortho_upper_bound}

Here, we divert our attention to the case where $M>2$. Let $A_{i+}$ be the event of detecting $+\s_i$, and $A_{i-}$ be the event of detecting $-\s_i$. Then, the probability of a symbol error given $+\bm{s}_i$ was transmitted is given by
\begin{equation} \label{eq:P_e_biortho_HD_true}
  P_{e|\s_i} = \Pr \giventhat*{\bigcup_{\substack{j=0 \\j \ne i}}^{M-1}\left(A_{j+} \cup A_{j-}\right) \cup A_{i-}}{+\bm{s}_i \text{ sent}}.
\end{equation}

The right-hand side of \eqref{eq:P_e_biortho_HD_true} should be stated as the probability of one or more the indicated events happens. Then, taking note that for any pair of events $A$ and $B$, $\Pr[A\cup B]\le \Pr[A]+\Pr[B]$,  \eqref{eq:P_e_biortho_HD_true} implies
\begin{equation} \label{eq:P_e_biortho_HD_sum1}
  P_{e|\s_i}  \le \sum_{\substack{j=0 \\j \ne i}}^{M-1}\Pr \giventhat*{A_{j+}  \cup A_{j-}  \cup A_{i-}  }{+\bm{s}_i \text{ sent}}.
\end{equation}

Making use of the results in Section~\ref{sec:M4_special_case}, this result can be written as
\begin{equation} \label{eq:P_e_biortho_HD_sum2}
  P_{e|\s_i}  \le \sum_{\substack{j=0 \\j \ne i}}^{M-1}\left(1 - \prod_{k=0}^{1}\left[1 - Q\left(\sqrt{\frac{2\En\rho_{i,j,k}^2}{N_0}}\right)\right]\right)
\end{equation}
where
\begin{equation}\label{eqn:rhos1}
  \begin{aligned}
    \rho_{i,j,0} & =\frac{(\s_i - \s_j)^\T \s_j }{|\s_i - \s_j|}   \\
    \rho_{i,j,1} & = \frac{(\s_i + \s_j)^\T \s_j }{|\s_i + \s_j|}.
  \end{aligned}
\end{equation}

Using \eqref{eq:P_e_biortho_HD_sum2} and assuming that the data symbols $\pm\s_i$, for $i=0,1,\cdots,N-1$, are equally likely to be transmitted, one will find that the probability of a symbol error, $P_e$,  is upper bounded by
\begin{equation}\label{eq:P_e_bound}
  P_e    \le \frac{1}{M}\sum_{i=0}^{M-1} \sum_{\substack{j=0            \\j \ne i}}^{M-1}\left(1 - \prod_{k=0}^{1}\left[1 - Q\left(\sqrt{\frac{2\En\rho_{i,j,k}^2}{N_0}}\right)\right]\right).
\end{equation}

The simulation results presented in Section~\ref{sec:sims} reveal that this upper bound  becomes tight as the SNR increases. This observation may be explained as follows. As the SNR increases, the events referred to on the right-hand side of \eqref{eq:P_e_biortho_HD_true} become nearly non-overlapping, equivalently, it will be unlikely that any pair of these events happen simultaneously. In that case, the inequality in \eqref{eq:P_e_biortho_HD_sum1} will become closer to an equality.

\subsection{Upper Bound for Quasi-Orthogonal Signaling} \label{sec:ortho_upper_bound}
The desired bound here can be derived following those of the biorthogonal case with some minor modifications. First, for the case where $M=2$, the decision boundary that separates $\s_0$ and $\s_1$ is the bisect line between them. This is the line in the direction $\textbf{d}_1$ in Fig.~\ref{fig:biortho_vectors}(a). With this observation, the probability of a symbol error here is found to be
\begin{equation}
  P_e=Q\left(\sqrt{\frac{2\En\rho_0^2}{N_0}}\right).
\end{equation}

Next, following the same line of thoughts as those in Section~\ref{sec:biortho_upper_bound}, and removing $A_{j-}$ and $A_{i-}$ from the right-hand side \eqref{eq:P_e_biortho_HD_true} and subsequent equations, it is not hard to arrive at the following upper bound for probability of symbol error for the present case.
\begin{equation}\label{eq:P_e_bound_ortho}
  P_e    \le \frac{1}{M}\sum_{i=0}^{M-1} \sum_{\substack{j=0            \\j \ne i}}^{M-1}Q\left(\sqrt{\frac{2\En\rho_{i,j,0}^2}{N_0}}\right)
\end{equation}

\color{black}

\section{Discussions and Insights} \label{sec:discussion}
The results developed in the previous sections, besides providing mathematical equations and numerical analysis for evaluating the SER of multi-coding  methods, lead to some insights that are worth noting. In this section, we present a summary of such insights.

First, correlation among the code vectors is not necessarily destructive. A well-known case of this is the simplex signaling where the code vectors are equally correlated as in \eqref{eqn:SS}. Here, from \eqref{eq:equi_corr_pc}, we note that the probability of correct detection of each symbol increases as $\eta$ decreases. This probability is maximized, hence, the SER is minimized, when $\eta$ is set equal to its minimum possible choice $\eta= -\frac{1}{M-1}$. From the results that follow \eqref{eqn:W1}, we note that this choice of $\eta$ leads to the eigenvalues $\lambda_0=0$ and $\lambda_i=\frac{M}{M-1}$, for $i=1,2,\cdots,M-1$. Also, any $\eta<-\frac{1}{M-1}$ leads to a negative eigenvalue, which then invalidates the condition that $\W$ is a correlation matrix. It is also worth noting that the choice $\eta= -\frac{1}{M-1}$ corresponds to the case that the eigenvalue spread/condition number of $\W$ is infinity. In the orthogonal case where $\eta=0$, on the other hand, the condition number of $\W$ is one. This observation leads us to the conclusion that the condition number of correlation matrix of the code vectors, i.e., $\W=\SM^{\rm H}\SM$, in a multi-coding signaling, in general, does not say much about its performance. In other words, the relationship between the correlation matrix $\W$ and the performance of the associated multi-coding system, in general, may be non-trivial.

\begin{figure}
  \centering
  \includegraphics[width=0.7\columnwidth]{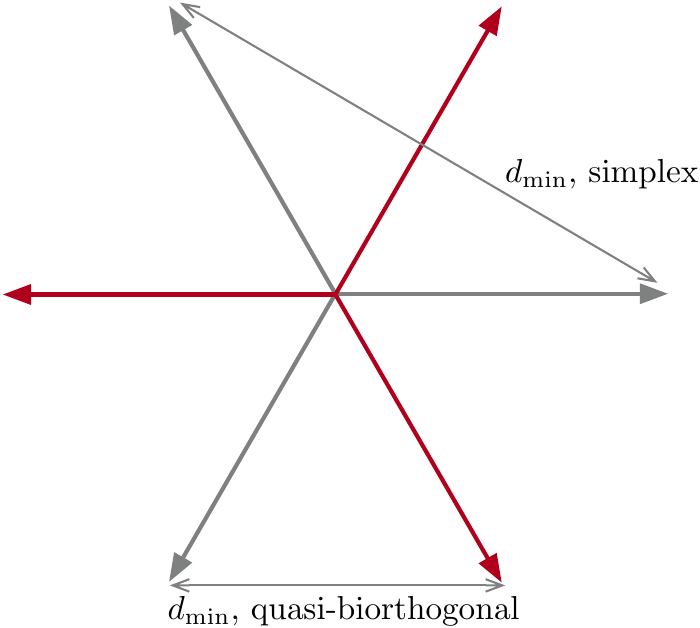}
  \caption{Simplex code vectors for the case when the number of code vectors are three. The negative of these vectors are also shown (in red). }
  \label{fig:simplexD3}
\end{figure}

Next, we take note that the above observations remain valid only for the case of quasi-orthogonal signaling. We may see a different picture when we consider the case of quasi-biorthogonal signaling. For instance, if we take a set of code vectors from the simplex signaling to construct a quasi-biorthogonal signaling system, the minimum distance between the new vector set decreases significantly. Fig.~\ref{fig:simplexD3} presents the set of code vectors of the simplex signaling for the case when the number of code vectors are three. In this case, the code vectors all fall in a plane of dimension two. The negative of these vectors, that should be included in a quasi-biorthogonal signaling system, are also shown. We have high-lighted the minimum distance, $d_\min$, between the vectors in both cases. In the case of biorthogonal signaling, the best performance is achieved by making use of a set of orthogonal codes, \cite{WeberC.1966Nstt}.

The numerical results presented in the next section reveal that the performance loss incurred due to any small deviation from a biorthogonal signaling remains negligible. Here, we provide a theoretical explanation to this observation. To this end, in the sequel, we focus on the upper bound in  \eqref{eq:P_e_bound_ortho} and explore its variation as the cross-correlation coefficients 
\begin{equation}\label{eqn:kappaij}
\kappa_{i,j}=\s_i^{\rm T}\s_j,\quad\mbox{for $0\le i,j\le M-1$ and $i\ne j$},
\end{equation}
deviate from zero.

We first take note that the upper bound in \eqref{eq:P_e_bound_ortho} can be rearranged as
\begin{equation}\label{eqn:Pub1}
P_{\rm ub}= \frac{2}{M}\sum_{i=0}^{M-1} \sum_{j=i+1}^{M-1}\left(1 - \prod_{k=0}^{1} \Phi\left(\sqrt{\frac{2\En\rho_{i,j,k}^2}{N_0}}\right)\right)
\end{equation}
where $\Phi(x)=1-Q(x)$ is the cumulative distribution function (CDF) of the standard normal function, and we have noted that $\rho_{i,j,k}=\rho_{j,i,k}$. Next, making use of \eqref{eqn:kappaij} 
and recalling that the code vectors $\s_i$ are unit length, it is straightforward to show that the equations in \eqref{eqn:rhos1} can be rearranged as
\begin{equation}\label{eqn:rhos2}
  \begin{aligned}
 \rho_{i,j,0} & =-\sqrt{\frac{1 - \kappa_{i,j}}{2}} \\
  \rho_{i,j,1} & = \sqrt{\frac{1 + \kappa_{i,j}}{2}}.
  \end{aligned}
\end{equation}
Substituting these in \eqref{eqn:Pub1}, it may be rearranged as in \eqref{eq:P_e_bound_simp_r}, shown at the top of the next page. In \eqref{eq:P_e_bound_simp_r}, it is explicitly shown that the upper bound  $P_{\rm ub}$ is a function of the coefficient vector 
\begin{equation}
  \bm\kappa = \begin{bmatrix}
    \bm\kappa_0^\T & \bm\kappa_1^\T & \cdots & \bm\kappa_{M-2}^\T
  \end{bmatrix}^\T, 
  \end{equation}
where   $  \bm\kappa_i = \begin{bmatrix}
    \kappa_{i,i+1} & \kappa_{i,i + 2} & \cdots & \kappa_{i, M-1}
  \end{bmatrix}^\T$. We are interested in variation of the upper bound $P_{\rm ub}(\bm\kappa)$ as $\bm\kappa$ deviates from zero.  This can be best done by looking at the Taylor series expansion of $P_{\rm ub}(\bm\kappa)$.

\begin{table*}
  \normalsize \begin{equation} \label{eq:P_e_bound_simp_r}
   P_{\rm ub}(\bm\kappa) =  M-1  -\frac{2}{M}\sum_{i=0}^{M-1} \sum_{j=i+1}^{M-1}\Phi\left(\sqrt{\frac{\En}{N_0}(1-\kappa_{i,j})}\right) 
  \Phi\left(\sqrt{\frac{\En}{N_0}(1+\kappa_{i,j})}\right).
  \end{equation}
  \hrulefill
\end{table*}

The Taylor series of $P_{\rm ub}(\bm\kappa)$ at $\bm\kappa=\bm{0}$ is written as
\begin{equation}\label{eqn:TS1}
P_{\rm ub}(\bm\kappa)=P_{\rm ub}(\bm{0})+\bm\kappa^{\rm T}\left(\nabla_{\bm\kappa}P_{\rm ub}\right)+\frac{1}{2!}\bm\kappa^{\rm T}\left(\nabla^2_{\bm\kappa}P_{\rm ub}\right)\bm\kappa+\cdots
\end{equation}
where the gradient $\nabla_{\bm\kappa}P_{\rm ub}$ and the Hessian matrix $\nabla^2_{\bm\kappa}P_{\rm ub}$ are evaluated at $\bm\kappa=\bm{0}$.

When the elements $\bm\kappa$ are small, the terms of orders greater than two in \eqref{eqn:TS1} will be negligible, hence, may be ignored. It is also straightforward  to evaluate the gradient vector $\nabla_{\bm\kappa}P_{\rm ub}$ and the Hessian matrix $\nabla^2_{\bm\kappa}P_{\rm ub}$ and evaluate their values at $\bm\kappa=\bm{0}$. These lead to
\begin{equation}
\left.\nabla_{\bm\kappa}P_{\rm ub}(\bm\kappa)\right|_{\bm\kappa=\bm 0}=\bm 0
\end{equation}
and
\begin{equation}
\left.\nabla^2_{\bm\kappa}P_{\rm ub}(\bm\kappa)\right|_{\bm\kappa=\bm 0}=\alpha\bf I 
\end{equation}
where
\begin{align} \label{eq:alpha_def}
\alpha=&\frac{1}{M}\sqrt{\frac{\En}{N_0}}\Biggl[\sqrt{\frac{\En}{N_0}}f^2\left(\sqrt{\frac{\En}{N_0}}\right)  \nonumber\\   
&+f\left(\sqrt{\frac{\En}{N_0}}\right)\Phi\left(\sqrt{\frac{\En}{N_0}}\right) \Biggl( \frac{\En}{N_0}  + 1 \Biggr) \Biggr].
\end{align}
Using the above results in \eqref{eqn:TS1}, it reduces to
\begin{equation}\label{eqn:TS2}
P_{\rm ub}(\bm\kappa)=P_{\rm ub}(\bm{0})+\alpha\bm\kappa^{\rm T}\bm\kappa.
\end{equation}

In Appendix~\ref{app:ts_ratio}, we evaluate the ratio of the coefficient $\alpha$ over $P_{\rm ub}(\bm{0})$ and find that it can be approximated as
\begin{equation}\label{eqn:TSratio}
\frac{\alpha}{P_{\rm ub}(\bm 0)}\approx \frac{1}{2M^2}\cdot \left(\frac{\En}{N_0}\right)^2.
\end{equation}
This result clearly shows that for large choices of the code size $M$ and moderate values of the SNR $\En/N_0$, $\alpha$ may remain significantly smaller than $P_{\rm ub}(\bm{0})$. Furthermore, we take note that for moderate to high values of $\En/N_0$, $P_{\rm ub}(\bm{0})$ is a good approximation to the symbol probability of the orthogonal case. Hence, one may argue, small deviations of $\bm\kappa$ from $\bm 0$, where $\bm\kappa^{\rm T}\bm\kappa$ remains smaller than one, may incur only a small degradation in performance, when compared to its orthogonal counterpart.

\color{black}

\section{Numerical Results} \label{sec:sims}
{\color{black} In this section, we demonstrate the analytical findings of the previous sections by evaluating the error probabilities for a few choices of correlated codes. \color{black}} To this end, we present a set of numerical results that compare the symbol error rates (SERs) of the quasi-orthogonal and quasi-biorthogonal signaling methods against their respective orthogonal and biorthogonal counterparts. Also, through numerical results, we examine the tightness of the upper bounds that were derived in Section~\ref{sec:UpperBound}. {\color{black} We explore codes generated through a few methods. These include codes generated from uniformly distributed random matrices with a maximum level of cross-correlation as well as the practical codes presented in \cite{ofmt_paper}. \color{black}}

Here, to generate a set of non-orthogonal codes, we begin with generating a random correlation matrix $\bf W$ with diagonal elements of unity. For $\bf W$ to be a valid correlation matrix, it should be symmetric, i.e., ${\bf W}^\T={\bf W}$, and all of its eigenvalues should be non-negative. To satisfy these conditions, first, the off diagonal elements in the upper triangular part of $\bf R$ are chosen to be random and independent taken from a uniform distribution in the interval $[-\rho_{\max},+\rho_{\max} ]$. These elements are then copied to the lower triangular part of $\bf W$ so that to satisfy the symmetry condition  ${\bf W}^\T={\bf W}$. Finally, to make sure that $\bf W$ is a valid correlation matrix, its eigenvalues are examined to be all non-negative. In case the latter condition is not satisfied, another random choice of $\bf W$ is generated until a proper choice $\bf W$ is obtained.

Once we have found a proper correlation matrix $\bf W$, we find the matrix $\SM$ which satisfies the equality $\SM^\T\SM={\bf W}$, hence, columns of $\SM$ are the desired multi-code symbol vectors. A matrix $\SM$ that satisfies the equality $\SM^\T\SM={\bf W}$ is often referred to as a square-root of $\bf W$. The common method of finding the square-root of a non-negative definite matrix $\bf W$ is to first perform the factorization ${\bf W}={\bf V}\bm\Lambda{\bf V}^\T$, where columns of $\bf V$ are eigenvectors of $\bf W$ and $\bm\Lambda$ is a diagonal matrix with the eigenvalues of $\bf W$ at its diagonal elements, and then set $\SM={\bf V}\bm\Lambda^{\frac 12}{\bf V}^\T$.

\begin{figure}
  \centering
  \includegraphics{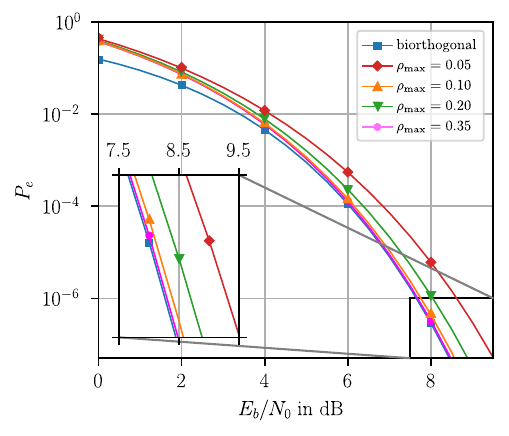}
  \caption{\color{black}SER, $P_e$, results for a range of non-orthogonal multi-codes from a quasi-biorthogonal set. The result arising from a set of biorthogonal codes is included for reference.\color{black}}
  \label{fig:ortho_cmp_curves}
\end{figure}

{\color{black}Fig.~\ref{fig:ortho_cmp_curves} presents the SER results of quasi-biorthogonal coding for the case where $M=16$. A few choices of the parameter $\rho_\max$ are examined. The SER plots, here, are found using the Monte Carlo integration techniques developed in Section~\ref{sec:mc_evaluation}. The SER plot of the biorthogonal coding is also presented as a bench-mark. As one would expect from the analysis in Section~\ref{sec:discussion}, for small choices of  $\rho_\max<0.1$, the non-orthogonality of multi-code vectors incurs very little loss in performance. The performance loss is about $0.4$~dB when $\rho_\max=0.2$, and is slightly above $1$~dB when $\rho_\max$ is set to a high value $0.35$. Similar results to those in Fig.~\ref{fig:ortho_cmp_curves} are obtained when the same code sets are applied to a quasi-orthogonal signaling. The SER curves for the orthogonal case will be shifted to the right by about $0.5$~dB. \color{black}}

\begin{figure}
  \centering
  \includegraphics{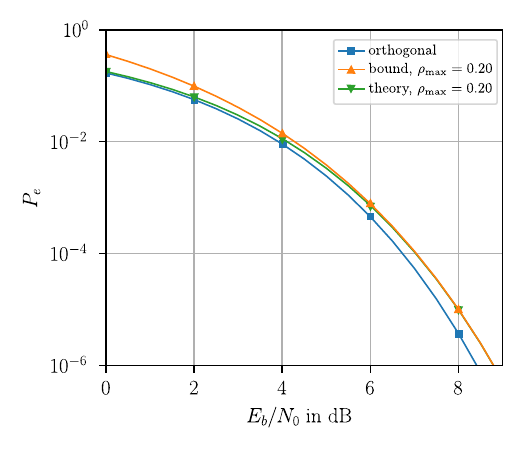}
  \caption{{\color{black}SER, $P_e$, results for non-orthogonal multi-codes with $\rho_{\max} = 0.2$. The result arising from a set of orthogonal codes is included for reference. The upper bound given by \eqref{eq:P_e_bound_ortho} is also presented.\color{black}}}
  \label{fig:bnd_curve}
\end{figure}

Fig.~\ref{fig:bnd_curve} shows the upper bound derived in Section~\ref{sec:ortho_upper_bound} for the quasi-orthogonal signaling scheme with $M=16$ and $\rho_{\max} = 0.2$. As can be seen, the upper bound becomes very tight at the symbol error rates of practical interest ($P_e<10^{-3}$).

\begin{figure}
  \centering
  \includegraphics{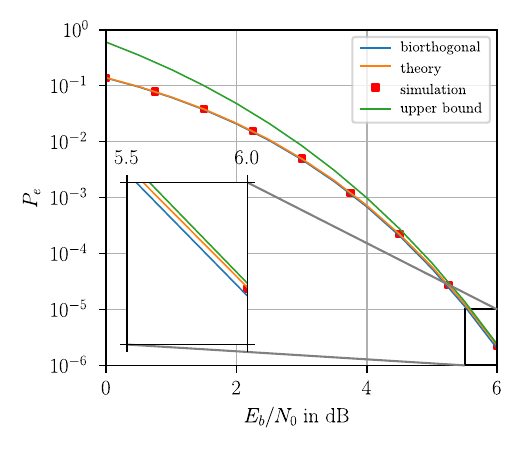}
  \caption{{\color{black}SER, $P_e$, results for practical non-orthogonal multi-codes where the average cross-correlation is $0.0224$ and the maximum cross-correlation is $0.0625$. \color{black}}}
  \label{fig:ofmt_curves}
\end{figure}

{\color{black}To demonstrate the error probability analysis for a set of practical codes, we choose the quasi-biorthogonal codes presented in \cite{ofmt_paper}. Here the code elements are chosen from the set $\{ \pm 1 \}$ and normalized so that the code vectors are unit length. Code vectors are then formed as circular shifts of $\s_0$, where $\s_0$ is the output of an optimization algorithm that attempts to minimize the cross-correlations among the circularly shifted multi-codes. Due to the restrictions on the code elements, the code vectors will not form an orthogonal set, nor be equi-correlated.  Fig.~\ref{fig:ofmt_curves} compares the theoretical error rate curve and a simulated error rate curve for a code set of this type of size where $L=M=128$. For this particular code set, the average cross-correlation of code vectors is $0.0224$ and the maximum cross-correlation is $0.0625$. Additionally, the upper bound and orthogonal theory curves are shown for comparison. As seen, our theoretical result is in perfect agreement with the simulation results and loss incurred due to the non-orthogonality of the code vectors is minimal in this design.  Also, as one would expect, the upper bound approaches the quasi-biorthogonal results as the SNR increases. \color{black}}

\section{Conclusion} \label{sec:conclusion}
A set of compact equations for symbol error probabilities of a general class of multi-coding systems, where the correlation among different code vectors can be any arbitrary value, were derived.  The derivation includes both quasi-orthogonal and quasi-biorthogonal signaling methods. The use of Monte Carlo integration was introduced for numerical calculation of the SERs of the studied cases. It was noted that this numerical method allows accurate evaluations of SERs down to the values as low as $10^{-6}$ with an affordable computational complexity. In addition, we developed equations for tight upper bounds of SERs of interest. These upper bounds provide accurate estimates of the SERs in the cases where their numerical calculation may be infeasible. {\color{black}Through a Taylor series expansion of the upper bound, we found analytical justification and explanation for the small loss in error probability for quasi-biorthogonal coding schemes.  \color{black}}

\section*{Acknowledgements}
This research made use of Idaho National Laboratory's High Performance Computing systems located at the Collaborative Computing Center and supported by the Office of Nuclear Energy of the U.S. Department of Energy and the Nuclear Science User Facilities under Contract No. DE-AC07-05ID14517.

\thanks{This manuscript has in part been authored by Battelle Energy Alliance, LLC under Contract No. DE-AC07-05ID14517 with the U.S. Department of Energy. The United States Government retains and the publisher, by accepting the paper for publication, acknowledges that the United States Government retains a nonexclusive, paid-up, irrevocable, world-wide license to publish or reproduce the published form of this manuscript, or allow others to do so, for United States Government purposes. STI Number: INL/JOU-24-77579.}

\appendices
{\color{black}
\section{Evaluation of $\alpha/P_{\rm ub}(\bm 0)$} \label{app:ts_ratio}
Evaluated at the biorthogonal code set, where $\kappa_{i,j}=0$,
\eqref{eq:P_e_bound_simp_r} reduces to
\begin{equation} \label{eq:upper_bnd_ortho}
  P_{\rm ub}(\bm 0) = (M - 1)Q\left(\sqrt{\frac{\En}{N_0}}\right)\left[2 - Q\left( \sqrt{\frac{\En}{N_0}}\right)\right].
\end{equation}
From \cite[pg. 42]{ProakisJohnG2008Dc}, we recall that
\begin{equation} \label{eq:q_func_bound}
  Q(x) < \frac{1}{x}f(x).
\end{equation}
Substituting \eqref{eq:q_func_bound} into \eqref{eq:upper_bnd_ortho}, we get
\begin{align} \label{eq:simp_ub}
  P_{\rm ub}(\bm 0) < (M - 1)&\left[\frac{2}{\sqrt{\frac{\En}{N_0}}}f\left(\sqrt{\frac{\En}{N_0}}\right) \right. \nonumber \\ 
  &\quad \left. -\frac{1}{\frac{\En}{N_0}}f^2\left(\sqrt{\frac{\En}{N_0}}\right)\right].
\end{align}
At moderate to high SNR values, the bound in \eqref{eq:q_func_bound} becomes tight and the second term in \eqref{eq:simp_ub} becomes insignificant, hence, one can say
\begin{align} \label{eq:high_snr_approx}
  P_{\rm ub}(\bm 0) \approx \frac{2(M - 1)}{\sqrt{\frac{\En}{N_0}}}f\left(\sqrt{\frac{\En}{N_0}}\right).
\end{align}
In addition, at moderate to high SNR values, where $\En/N_0\gg 1$ and $f\left(\sqrt{\frac{\En}{N_0}}\right)\ll 1$, 
 \eqref{eq:alpha_def} can be simplified as
\begin{align} \label{eq:second_deriv_high_snr}
\alpha \approx &\frac{1}{M}\left(\frac{\En}{N_0}\right)^{\frac{3}{2}} f\left(\sqrt{\frac{\En}{N_0}}\right).
\end{align}
Making use of \eqref{eq:high_snr_approx} and \eqref{eq:second_deriv_high_snr}, and taking note that for typical choices of interest $M-1\approx M$, one arrives at \eqref{eqn:TSratio}.
 \color{black}}

\bibliographystyle{IEEEtran}

\bibliography{IEEEabrv,ref}
\balance
\begin{IEEEbiography}[{\includegraphics[width=1in,height=1.25in,clip,keepaspectratio]{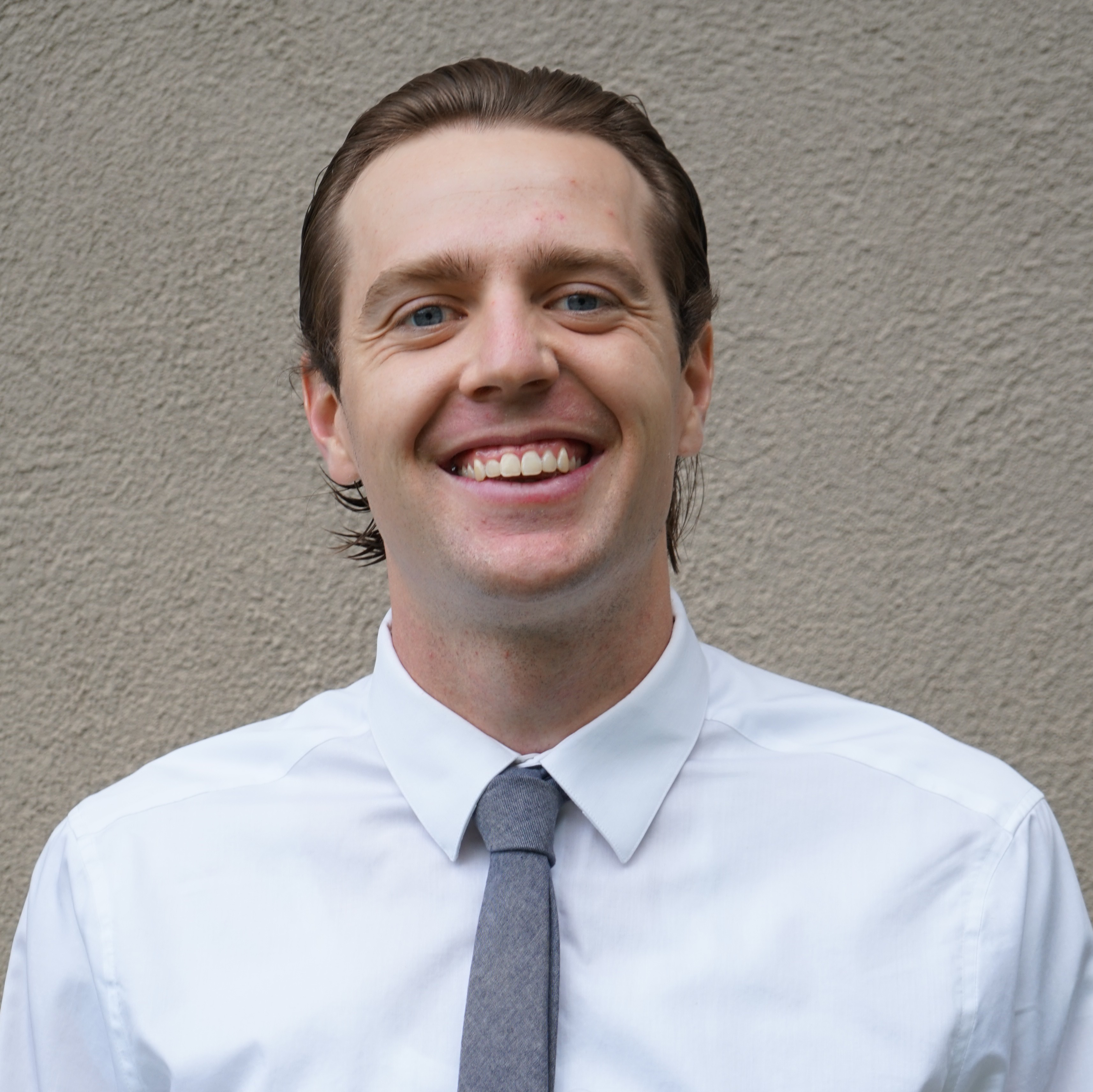}}]{Brian Nelson } (Member, IEEE) became a member of IEEE in 2025.  He received the B.S. Summa Cum Laude in computer engineering from Brigham Young University in Provo, UT, USA in 2021 and an M.S. degree in electrical engineering from the University of Utah in Salt Lake City, UT, USA in 2024. He is currently pursuing a PhD degree in electrical engineering from the University of Utah. His PhD research has focused on spread spectrum and ultra-wideband communications.

  He was an engineer in the modem group at L3Harris Technologies from 2021 to 2022. Since 2022, he  has worked as a researcher in the Idaho National Laboratory Wireless Research Department in Salt Lake City, UT, USA.
\end{IEEEbiography}

\begin{IEEEbiography}[{\includegraphics[width=1in,height=1.25in,clip,keepaspectratio]{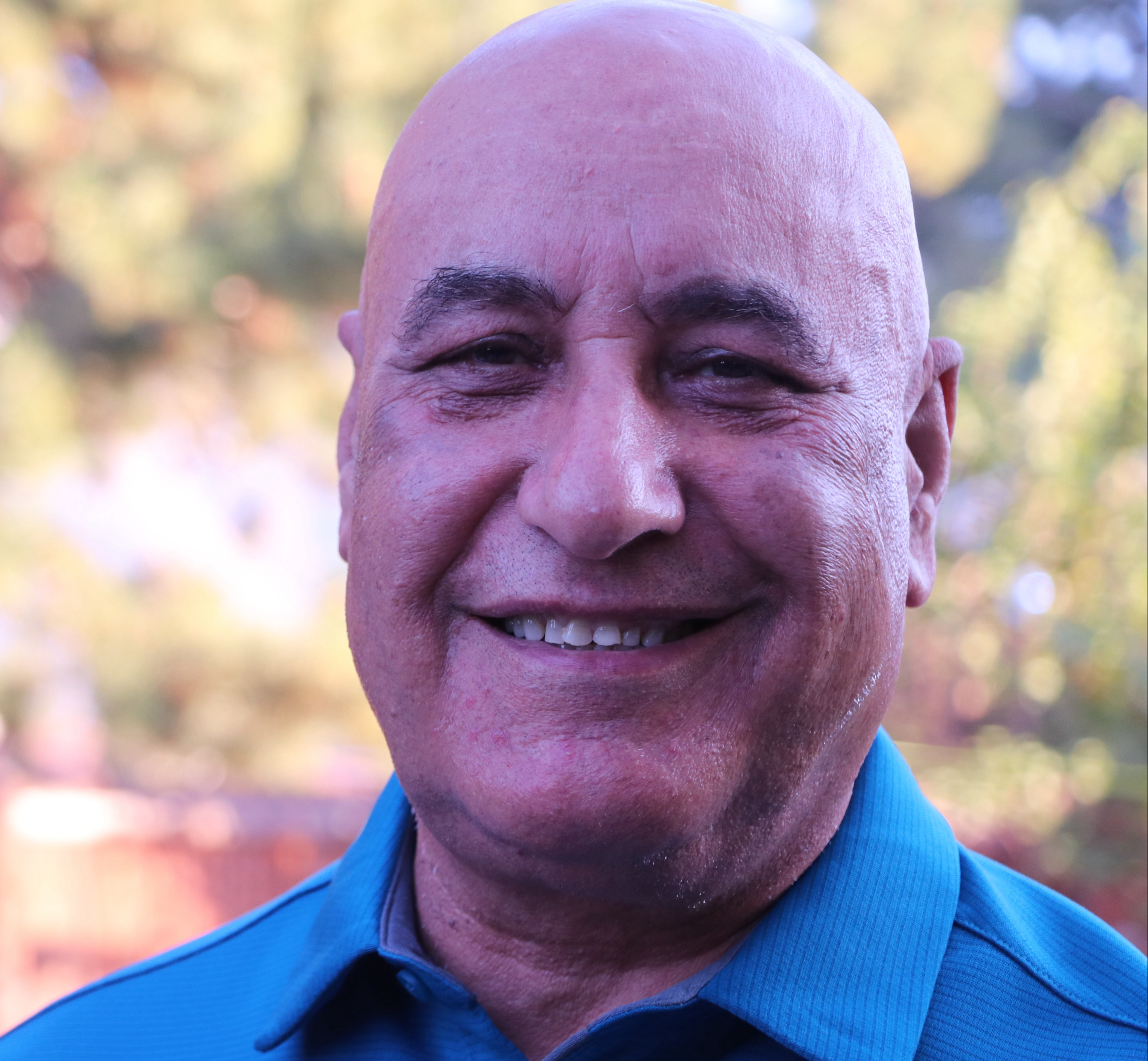}}]{Behrouz Farhang-Boroujeny }  (M'84, SM'90, Life SM, 2017) received the B.Sc. degree in electrical engineering from Teheran University, Iran, in 1976, the M.Eng. degree from University of Wales Institute of Science and Technology, UK, in 1977, and the Ph.D. degree from Imperial College, University of London, UK, in 1981.

  From 1981 to 1989 he was with the Isfahan University of Technology, Isfahan, Iran. From 1989 to 2000 he was with the National University of Singapore. Since August 2000, he has been with the University of Utah.

  He is an expert in the general area of signal processing. His current scientific interests are adaptive filters, multicarrier communications, detection techniques for space-time coded systems, and cognitive radios. In the past, he has worked and has made significant contribution to areas of adaptive filters theory, acoustic echo cancellation, magnetic/optical recoding, and digital subscriber line technologies. He is the author of the books ``Adaptive Filters: theory and applications'', John Wiley \& Sons, 1998, and “Signal Processing Techniques for Software Radios”, self published at Lulu publishing house, 2009 and 2010 (second edition).
  Dr. Farhang-Boroujeny received the UNESCO Regional Office of Science and Technology for South and Central Asia Young Scientists Award in 1987. He served as an associate editor of IEEE Trans. on Signal Processing from July 2002 to July 2005, and as an associate editor of IEEE Signal Processing Letters from April 2008 to March 2010. He has also been involved in various IEEE activities, including the chairmanship of the Signal Processing/Communications chapter of IEEE of Utah in 2004 and 2005.
\end{IEEEbiography}

\end{document}